\documentclass[aip,amsmath,amssymb,reprint]{revtex4-1}

\usepackage{amsmath,bbm,amsfonts,amsthm,graphicx}
\usepackage{epstopdf}
\usepackage{dcolumn}
\usepackage{bm}
\usepackage[version=4]{mhchem}
\usepackage[utf8]{inputenc}
\usepackage[T1]{fontenc}
\usepackage{mathptmx}
\usepackage[colorlinks=true,breaklinks=true,linkcolor=blue,citecolor=blue,urlcolor=blue]{hyperref}

\newcommand{\multipole}{M}
\newcommand*{\citen}[1]{%
  \begingroup
    \romannumeral-`\x 
    \setcitestyle{numbers}%
    \cite{#1}%
  \endgroup   
}

\begin{document}

\title{Quantum computation of dominant products in lithium-sulfur batteries}

\author{Julia E. Rice}
\thanks{corresponding author, jrice@us.ibm.com}
\affiliation{IBM Quantum, Almaden Research Center, San Jose, CA 95120, USA}

\author{Tanvi P. Gujarati}
\affiliation{IBM Quantum, Almaden Research Center, San Jose, CA 95120, USA}

\author{Mario Motta}
\affiliation{IBM Quantum, Almaden Research Center, San Jose, CA 95120, USA}

\author{Tyler Y. Takeshita}
\affiliation{Mercedes Benz Research and Development North America, Sunnyvale, CA 94085, USA}

\author{Eunseok Lee}
\affiliation{Mercedes Benz Research and Development North America, Sunnyvale, CA 94085, USA}

\author{Joseph A. Latone}
\affiliation{IBM Quantum, Almaden Research Center, San Jose, CA 95120, USA}

\author{Jeannette M. Garcia}
\affiliation{IBM Quantum, Almaden Research Center, San Jose, CA 95120, USA}

\begin{abstract}
Quantum chemistry simulations of some industrially relevant molecules are reported, 
employing variational quantum algorithms for near-term quantum devices.
The energies and dipole moments are calculated along the dissociation curves for 
lithium hydride (\ce{LiH}), hydrogen sulfide (\ce{H2S}), lithium hydrogen sulfide 
(\ce{LiSH}) and lithium sulfide (\ce{Li2S}). 
In all cases we focus on the breaking of a 
single bond, to obtain information about the 
stability of the molecular species being 
investigated.
We calculate energies and a variety of 
electrostatic properties of these molecules 
using classical simulators of quantum devices, with 
up to 21 qubits for lithium sulfide. Moreover, we 
calculate the ground-state energy and dipole 
moment along the dissociation pathway of 
\ce{LiH} using IBM quantum devices. 
This is the first example, to the best of our knowledge, of dipole moment calculations 
being performed on quantum hardware.
\end{abstract}

\maketitle

\section*{Introduction}

\begin{figure}[b!]
\includegraphics[width=0.65\columnwidth]{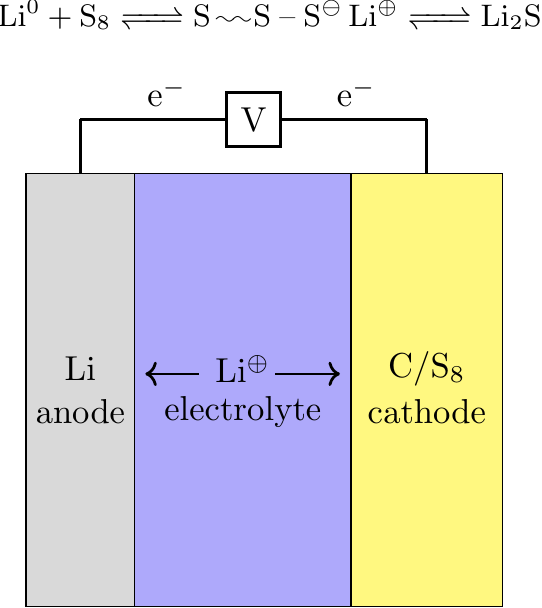} 
\caption{Schematic representation of a lithium-sulfur battery. During discharge, 
elemental sulfur is reduced into sulfur salts containing chains of varying lengths, 
ultimately resulting in lithium sulfide production. During charge, the chemistry 
is reversed. The mechanism for lithium sulfide production is proposed to occur 
through a two-electron process.}
\label{figure:1}
\end{figure}

Lithium-sulfur batteries are a promising next-generation battery technology with a high theoretical capacity of up to $\sim$1675 mAh/g and a high theoretical energy density of 
$\sim$2600 Wh/kg (\ce{Li}-ion theoretical energy density is $\sim$350-500 Wh/kg)
\cite{gibot2008room,bruce2012li,wild2015lithium,hagen2015lithium,zheng2016reduction,fang2017more}. 
The reversible battery functioning is based on the electrochemistry between lithium metal (\ce{Li^0}) and elemental sulfur (\ce{S8}) 
to form lithium sulfide (\ce{Li2S}) as the thermodynamic product, illustrated in Figure \ref{figure:1}. 
The reduction of sulfur is known to proceed on discharge in the battery, through a series of intermediate lithium polysulfide salts, 
some of which are soluble in the electrolyte and can diffuse to the lithium metal anode \cite{yamin1988lithium,mikhaylik2004polysulfide,barchasz2012lithium,cuisinier2014unique,pascal2014x}. The solubility of various intermediates provides 
unique challenges for the realization of lithium-sulfur batteries and limits the potential use of lithium-sulfur as a practical alternative to state-of-the-art lithium-ion \cite{chung2018progress,li2019comprehensive}. 

In order to harness control over the battery technology,
the mechanism for production and identification of reactive intermediates must be validated. 
However, the reaction mechanism for sulfur reduction in the battery environment is highly complex and debated in the field. 
Intermediates are difficult (if not impossible) to characterize during operation of the battery and competing chemical (vs electrochemical) 
pathways are also active, which confounds the mixture of products produced. Generally, products are assumed to be produced through 
either one electron or two electron processes, which generate radicals and \ce{Li}-\ce{S} salts, in which the sulfur anion consists of 
1-8 sulfur atoms. Four electron processes have also been considered but are unlikely as there is little experimental 
evidence and for entropic reasons \cite{wujcik2018determination}.

The molecular electron-density distribution and, in particular, the resulting dipole moment, are critical for understanding a variety 
of phenomena occurring in batteries.  In general, molecules with high polarity can easily attract or repel valence electrons from 
other compounds and generate reactions through electron transfer. The dipole moment of a molecule also determines its response 
to an external electric field.
Accurate computation of energetics and dipole moments of molecules is thus a problem with deep conceptual importance, and 
significant applicability to the chemistry of LiS batteries. Achieving this goal requires solving the Schr\"odinger equation for the 
molecules of interest, a problem that is known to be exponentially expensive for classical computers 
unless approximation schemes are introduced.

Quantum computing is an alternative and complementary mode of attack of mathematical problems, that has significant potential to provide advantage over conventional computing in a number of areas, including simulation of Hamiltonian dynamics \cite{aspuru2005simulated,georgescu2014quantum,kassal2011simulating,cao2019quantum}.
A number of heuristics to provide approximate but highly accurate solutions to the Schr\"odinger equation have been proposed, 
in particular the Variational Quantum Eigensolver (VQE) \cite{peruzzo2014variational,kandala2017hardware,ollitrault2019quantum}. 
Over the last years, researchers have demonstrated the use and accuracy of VQE in investigations of a variety of molecules \cite{peruzzo2014variational,omalley2016scalable,kandala2017hardware,colless2018computation,gao2019computational,bauer2020quantum,mcardle2020quantum,grimsley2019adaptive,bian2019quantum,nam2020ground}.

Motivated by these results, and by the importance of computing energies and electrostatic properties, in this work we assess 
the performance of quantum algorithms in determining ground state energies and dipole moments along bond stretching for 
\ce{LiH}, \ce{H2S}, \ce{LiSH} and \ce{Li2S}, using classical simulators of quantum devices.
The calculations presented here are important for the development of quantum computing: to the best of our knowledge, we 
present the first quantum hardware results of dipole moments, focusing on the \ce{LiH} molecule. Our results immediately generalize to expectation values of $k$-body operators in materials, with significant implications for the ability to study reactivity and electrostatics in 
batteries by quantum algorithms.

\section*{Methodology}

\subsection{Working equations}

The overall strategy for the calculations performed in this work involved initial pre-processing by classical quantum chemistry codes on conventional computers, 
to generate optimized Hartree-Fock orbitals and matrix elements of the Hamiltonian, prior to performing computations with quantum simulators or devices.
The restricted Hartree-Fock (RHF) singlet state has been chosen as the initial state for all of the calculations described here, since experience has indicated 
this state as a good choice for a variety of chemical problems \cite{romero2018strategies}. 
Restricted coupled-cluster with singles and doubles (CCSD) and full configuration interaction (FCI)
calculations were performed using Psi4 \cite{turney2012psi}, 
at STO-3G level of theory, using the frozen-core approximation for correlated calculations. 
The choice of the minimal basis is motivated by the fact that only a few molecular orbitals can be described on contemporary quantum hardware, 
since the number of available qubits is still relatively small, devices are noisy, and full quantum error correction techniques are not yet available. 
Additional details for the studied molecules are listed in the Supplementary Material.

Having selected a set of single-electron orbitals for each of the studied species, VQE computations were performed with quantum simulators and devices. 
{Achieving this goal requires first defining a Hilbert space
$\mathcal{H}$ spanned by $2n$ orthonormal one-electron wavefunctions $\{ \phi_p \otimes \chi_\sigma \}_{p\sigma}$, where $\{ \phi_p \}_{p=1}^n$ are a set of orthonormal spatial orbitals, here RHF  orbitals, and $\chi_\sigma$ are spin-$z$ eigenfunctions, with $\sigma = \uparrow, \downarrow$.

For each of the $2n$ spin-orbitals $\phi_p \otimes \chi_\sigma$, we then define a creation 
operator $\hat{c}^\dagger_{p\sigma}$, and construct the fermionic Fock space $\mathcal{F}$.
Now, $\mathcal{F}$ is of course a $2^{2n}$-dimensional Hilbert space, which is
very naturally mapped \cite{jordan1993paulische,bravyi2002fermionic,seeley2012bravyi} onto 
the Hilbert space of $m=2n$ qubits. 
An important example of such a mapping is the Jordan-Wigner transformation
\begin{equation}
\begin{split}
\hat{c}^\dagger_{p\uparrow} & \mapsto \frac{X_p-iY_p}{2} 
\, Z^{p-1}_0 \;, \\
\hat{c}^\dagger_{p\downarrow} &\mapsto \frac{X_{n+p}-iY_{n+p}}{2} \, Z^{n+p-1}_0 \;, \\
\hat{c}_{p\uparrow} & \mapsto \frac{X_p+iY_p}{2} 
\, Z^{p-1}_0 \;, \\
\hat{c}_{p\downarrow} &\mapsto \frac{X_{n+p}+iY_{n+p}}{2} \, Z^{n+p-1}_0 \;, \\
\end{split}
\end{equation}
with $Z^{r}_s = Z_r Z_{r-1} \dots Z_{s+1} Z_s$. Of course there exist several alternatives, most notably the parity and Bravyi-Kitaev encodings, that can be combined with ``tapering" techniques to 
reduce the number $m$ of qubits leveraging conservation of particle number modulo 2 and the presence of $\mathbb{Z}_2$ symmetries of molecular orbitals \cite{bravyi2002fermionic,setia2018bravyi,seeley2012bravyi,verstraete2005mapping,bravyi2017tapering,setia2019superfast,steudtner2017lowering}.
When fermionic wavefunctions can be mapped onto $m$-qubit wavefunctions, fermionic $k$-body operators are mapped onto 
linear combination of Pauli operators acting on $m$ qubits,
\begin{equation}
\sum_{ \substack{p_1 \dots p_k \\ q_1 \dots q_k}}
O^{p_1 \dots p_k}_{q_1 \dots q_k} \, 
\hat{E}^{q_1 \dots q_k}_{p_1 \dots p_k}
\mapsto
\sum_i c_i P_i
\;,
\end{equation}
where $P_i \in \{ \mathbbm{1}, X, Y, Z \}^{\otimes m}$
is an $m$-qubit Pauli operator, and
\begin{equation}
\hat{E}^{q_1 \dots q_k}_{p_1 \dots p_k} =
\sum_{\sigma_1 \dots \sigma_k}
\hat{c}^\dagger_{p_1 \sigma_1} \dots 
\hat{c}^\dagger_{p_k \sigma_k} \hat{c}^{\phantom{\dagger}}_{q_k \sigma_k}
\dots \hat{c}^{\phantom{\dagger}}_{q_1 \sigma_1}
\end{equation}
is a spin-summed $k$-body excitation operator.
For example, under Jordan-Wigner representation, one-body operators 
$\hat{O} = \sum_{pq} O^p_{q} \hat{E}^q_p$ transform onto
\begin{equation}
\begin{split}
&\hat{O} 
\mapsto
\sum_p O^p_{p} \frac{ \mathbbm{1} - Z_p }{2} +
\sum_p O^p_{p} \frac{ \mathbbm{1} - Z_{n+p} }{2} \\
&+
\sum_{p<q} O^p_{q} \frac{X_q Z^{q-1}_{p+1} X_p + Y_q Z^{q-1}_{p+1} Y_p}{2} \\
&+
\sum_{p<q} O^p_{q}
\frac{ X_{n+q} Z^{n+q-1}_{n+p+1} X_{n+p} + Y_{n+q} Z^{n+q-1}_{n+p+1} Y_{n+p} }{2}
\end{split}
\end{equation}
Once $k$-body operators are encoded onto qubit operators, they 
can of course be measured, using standard techniques, on a 
register of qubits prepared in a suitable wavefunction.
In this work, to produce an accurate approximation for the ground state of the system, we use the VQE algorithm, in which a 
parametrized quantum circuit $\hat{U}(\theta) | \Psi_0 \rangle$ 
is used to correlate an initial mean-field wavefunction $\Psi_0$, corresponding to the RHF state, and the best approximation
to the ground state is determined by numerically minimizing the
energy $E(\theta) = \langle \Psi_0 | \hat{U}(\theta)^\dagger \hat{H} \hat{U}(\theta) | \Psi_0 \rangle$.
Once the optimal parameters are found, properties can be evaluated as expectation values of suitable qubit operators over the VQE wavefunction.

Here, we focused on multipole moments, which we evaluate as expectation values of one-body operators, 
\begin{equation}
\multipole^{i_1 \dots i_n} = \multipole_{n}^{i_1 \dots i_n} - \langle \Psi | \hat{\multipole}_{e}^{i_1 \dots i_n} | \Psi \rangle
\end{equation}
given by a nuclear and an electronic contributions,
\begin{equation}
\begin{split}
\multipole_{n}^{i_1 \dots i_n} &= \sum_\alpha Z_\alpha R_\alpha^{i_1} \dots R_\alpha^{i_n} \\
\hat{\multipole}_{e}^{i_1 \dots i_n} &= \sum_{pq} \left(\multipole^{i_1 \dots i_n}\right)^p_q \hat{E}^q_p \\
\left(\multipole^{i_1 \dots i_n}\right)^p_q &= \int d^3{\bf{r}} \, r^{i_1} \dots r^{i_n} \, \phi_p({\bf{r}}) \phi_q({\bf{r}})
\end{split}
\end{equation}
respectively, where $i_1 \dots i_n = x,y,z$ are indices labeling Cartesian components, and $Z_\alpha$, $R_\alpha$ are the atomic numbers and the positions of the nuclei respectively.

From the quantities $\multipole^{i_1 \dots i_n}$, we easily obtain the dipole, quadrupole, octopole and hexadecapole moments in the traceless Buckingham expansion \cite{stone2013theory,buckingham1959molecular,buckingham1959direct,buckingham1967permanent,gray1976spherical}, respectively defined as 
\begin{widetext}
\begin{equation}
\begin{split}
\mu^a &= \multipole^a \\
\Theta^{ab} &= \frac{1}{2!} \Big[ 3 \multipole^{ab} - \multipole^{kk} \, \delta^{ab} \Big] \\
\Omega^{abc} &= \frac{1}{3!} \Big[ 15 \multipole^{abc} - 3 
\big(
\multipole^{akk} \, \delta^{bc} + 
\multipole^{kbk} \, \delta^{ac} +
\multipole^{kkc} \, \delta^{ab} \big) 
\Big] \\
\Phi^{abcd} &= \frac{1}{4!}  \Big[ 105 \multipole^{abcd} 
-15 \, \big( 
\multipole^{abkk} \, \delta^{cd} +
\multipole^{akck} \, \delta^{bd} + 
\multipole^{akkd} \, \delta^{bc} + 
\multipole^{kbck} \, \delta^{ad} +
\multipole^{kbkd} \, \delta^{ac} +
\multipole^{kkcd} \, \delta^{ab} \big) \\
&\phantom{aaaa}+ 3 \left( 
\multipole^{kkll} \, \delta^{ab} \, \delta^{cd} +
\multipole^{klkl} \, \delta^{ac} \, \delta^{bd} +
\multipole^{kllk} \, \delta^{ad} \, \delta^{bc}
\right)
\Big] \;,
\end{split}
\end{equation}
\end{widetext}
where repeated indices ($k,l = x,y,z$) are summed over.
Multipole moments are converted to spherical molecular moments 
as detailed in Table E1 of Ref~[\citen{stone2013theory}].

We evaluate the charge density and molecular electrostatic potential, respectively defined as
\begin{equation}
\rho({\bf{r}}) = \sum_{pq} \phi_p({\bf{r}}) \phi_q({\bf{r}}) 
\langle \Psi | \hat{E}^q_p | \Psi \rangle 
\end{equation}
and
\begin{equation}
\begin{split}
V({\bf{r}}) &= V_n({\bf{r}}) - V_e({\bf{r}}) \, , \\
V_n({\bf{r}}) &= \sum_\alpha \frac{Z_\alpha}{\|{\bf{r}}-{\bf{R}}_\alpha\|} \, , \\
V_e({\bf{r}}) &= \int d^3{\bf{r}}^\prime \, \frac{\rho({\bf{r}})}{\|{\bf{r}}-{\bf{r}}^\prime \|} = \sum_{pq} V_{pq} \, \langle \Psi | \hat{E}^q_p | \Psi \rangle
\end{split}
\end{equation}

In this work, we elected to evaluate multipole moments as expectation values of one-body operators. There exists of course an alternative approach, based on the definition of the dipole moment as the derivative of the ground-state energy with respect to an external electric field \cite{fitzgerald1986analytic,rice1986analytic,scheiner1987analytic,salter1989analytic,gauss1991analytic,gauss1991coupled,lodi2008new}.

Within such an approach, the dipole moment definition includes additional terms that depend on the derivatives of the Hartree-Fock orbitals with respect to the external electric field. 

The strategy of computing dipole moments as expectation values of one-body operators is simpler and more economical, as it requires $\mathcal{O}(n^3)$ rather than $\mathcal{O}(n^5)$ measurements of Pauli operators when the Jordan-Wigner encoding is used. Moreover, 
it does not require calculating energies within sub-milli-Hartree statistical accuracy, making it suitable for contemporary quantum 
hardware.

The implementation and hardware demonstration of gradient-based approaches is a valuable direction of research in the development 
and refinement of experiments on quantum hardware \cite{kassal2009quantum,mitarai2020theory,o2019calculating,sokolov2020microcanonical,parrish2019hybrid}, especially important in the
investigation of second-order derivatives, such as polarizabilities and shielding tensors.

We defined the charge density and electrostatic potential in terms of the spin-summed one-body density matrix
$\langle \Psi | \hat{E}^q_p | \Psi \rangle$ (1RDM) because,
unlike multipole moments, these quantities are evaluated on very large meshes of spatial points. Therefore, they are more conveniently computed by measuring and post-processing the 1RDM.
Multipole moments, on the other hand, are more naturally evaluated as expectation values of hermitian operators, without extracting and post-processing the 1RDM.
}

\subsection{Computational details}

We use IBM's open-source Python library for quantum computing, Qiskit \cite{aleksandrowicz2019qiskit}. 
Qiskit provides tools for various tasks such as creating quantum circuits, performing simulations, and computations on real hardware. 
It also contains an implementation of the VQE algorithm, 
a hybrid quantum-classical algorithm that uses both quantum and classical resources to solve the Schr\"{o}dinger equation and 
a classical exact eigensolver algorithm to compare results.

In the VQE algorithm, we take our wavefunction in the form of a quantum circuit, which is the unitary coupled cluster with singles 
and doubles (UCCSD) \cite{kutzelnigg1982quantum,kutzelnigg1983quantum,kutzelnigg1985quantum}, 
its quantum variant (q-UCCSD) as defined in Ref~[\citen{barkoutsos2018quantum}],
and the following $\mathrm{R_y}$ Ansatz,
\begin{equation}
| \Psi(\theta) \rangle = 
\prod_{k=1}^{n_r}
\left(
\prod_{i=0}^{m-1} R_{y,i}(\theta_{k,i})
\prod_{ij\in C} G_{ij}
\right)
\prod_{i=0}^{m-1} R_{y,i}(\theta_{0,i}) | \Psi_0 \rangle
\;,
\end{equation}
where $| \Psi_0 \rangle$ is an initial wavefunction (here, the restricted 
closed-shell Hartree-Fock state), $m$ is the number of qubits, $R_{y,i}(\theta) 
= \mbox{exp}(-i \theta Y_i/2)$ is a $Y$ rotation of an angle $\theta$ 
applied to qubit $i$, $G_{ij}$ a parameter-free two-qubit entangling 
gate (here, the CNOT gate) applied to a pair $(ij)$ of connected qubits 
(here, we chose linear connectivity, i.e. $(ij) \in C$ if and only if $j=i+1$),
and $n_r$ is an integer denoting the number of times a layer of entangling gates
followed by a layer of $Y$ rotations is repeated.

We then minimize the expectation value of the Hamiltonian with respect to the parameters of our circuit. The minimization is carried out through the classical optimization method, 
L$\_$BFGS$\_$B \cite{zhu1997algorithm,byrd1995limited,morales2011remark} on the simulator, 
and Simultaneous Perturbation Stochastic Approximation (SPSA) \cite{spall1998overview,spall2000adaptive} on the device. 
Once the VQE is complete, we obtain the optimized variational form and the estimate for the 
ground state energy. In addition, we measure the one-body operators corresponding to 
the components of the dipole, quadrupole, octopole and hexadecapole moments, and the electron density and electrostatic potential from the spin-summed one-particle density matrix $\langle \Psi | \hat{E}^q_p | \Psi \rangle$.

\begin{figure}[b!]
\includegraphics[width=0.85\columnwidth]{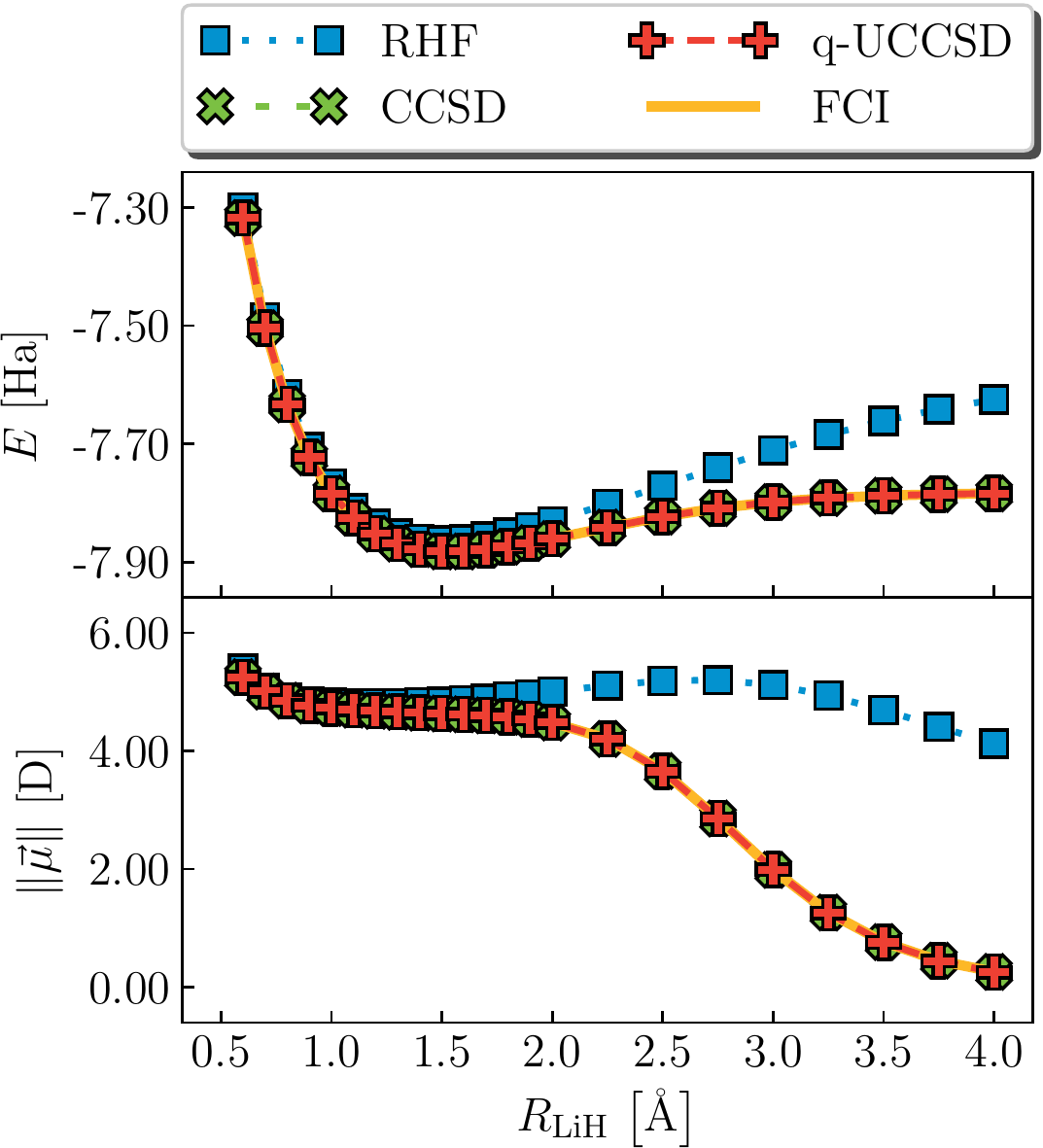}
\caption{
Calculations for lithium hydride, \ce{LiH}. 
Top: Dissociation curve (ground state energy in Hartrees) as a function of interatomic distance (in Angstroms) from RHF, CCSD, q-UCCSD and FCI. 
Bottom: Dipole moment (in Debye) as a function of interatomic distance (in Angstroms).
}
\label{figure:2}
\end{figure}

We chose molecular geometries to lie in the $xz$ plane of a suitable frame of reference, with the center of mass occupying the origin of such a frame of reference, and the heaviest element having $z<0$.

We ran our experiments on both the statevector and qasm simulators in Qiskit. 
We performed hardware experiments on 5 qubit devices available through IBM Quantum Experience each 
with Quantum Volume \cite{cross2019QV} of 32, namely, \textit{ibmq$\_$rome}, \textit{ibmq$\_$athens},
\textit{ibmq$\_$bogota} and \textit{ibmq$\_$santiago} \cite{IBMQDevices}. We employed readout-error
mitigation \cite{temme2017error,kandala2019error,bravyi2020mitigating} as implemented in Qiskit Ignis 
to correct measurement errors. 
We also used a simple noise extrapolation scheme using additional CNOT gates at the minimum energy 
VQE iterations to account for errors introduced during the expensive 2-qubit entangling operations 
as was shown in Refs~[\citen{dumitrescu2018cloud,stamatopoulos2019option}].

\section*{Results and Discussion}

\subsection{Quantum computation of lithium-sulfur molecules}

The stability of \ce{Li_2S} in batteries is related to the amount of energy needed to break one
\ce{Li-S} bond. So. in this study, we have studied the dissociation of one bond, rather than two 
bonds simultaneously.

\subsubsection {Lithium hydride $(\ce{LiH})$}

After removing the frozen core orbital (\ce{Li} 1s), the lithium hydride (\ce{LiH}) bond can be represented by molecular (spatial) orbitals (MOs) with 2s, 2p$_z$ character for lithium (since 
\ce{LiH} is placed along the $z$ axis) and an MO with 1s character for hydrogen, for a total of 
6 spin-orbitals. 
The ground state energy and dipole moment norm were calculated over a range of bond distances on the 
simulator and on the hardware through the  use of four qubits after tapering 
\cite{bravyi2017tapering} techniques were applied.

As seen in Figure \ref{figure:2}, results from CCSD and FCI coincide for both energy and dipole 
moment along bond stretching, since these methods are equivalent for two-electron systems. 
Also q-UCCSD values are essentially the same as FCI.
Despite its simplicity, \ce{LiH} has an interesting evolution in the dipole moment along the 
dissociation curve from a regime with ionic character 
(polar, short $R_{\ce{LiH}}$, $\| \vec{\mu} \| \simeq 4.6$ D) to one without polarity (large $R_{\ce{LiH}}$, $\| \vec{\mu} \| \simeq 0$). 
The large change in the dipole moment between 2.0 and 3.5 \AA $\,$ correlates with the pronounced deviation of the energy dissociation curve from that of the FCI in the first results reported 
on quantum hardware, see Figure 3 of Ref~[\citen{kandala2017hardware}]. Both of these observations 
are consistent with the fact that there is more entanglement/electron correlation across this 
bond length range, and thus higher circuit depth is needed.

\begin{figure*}[t!]
\includegraphics[width=0.6\textwidth]{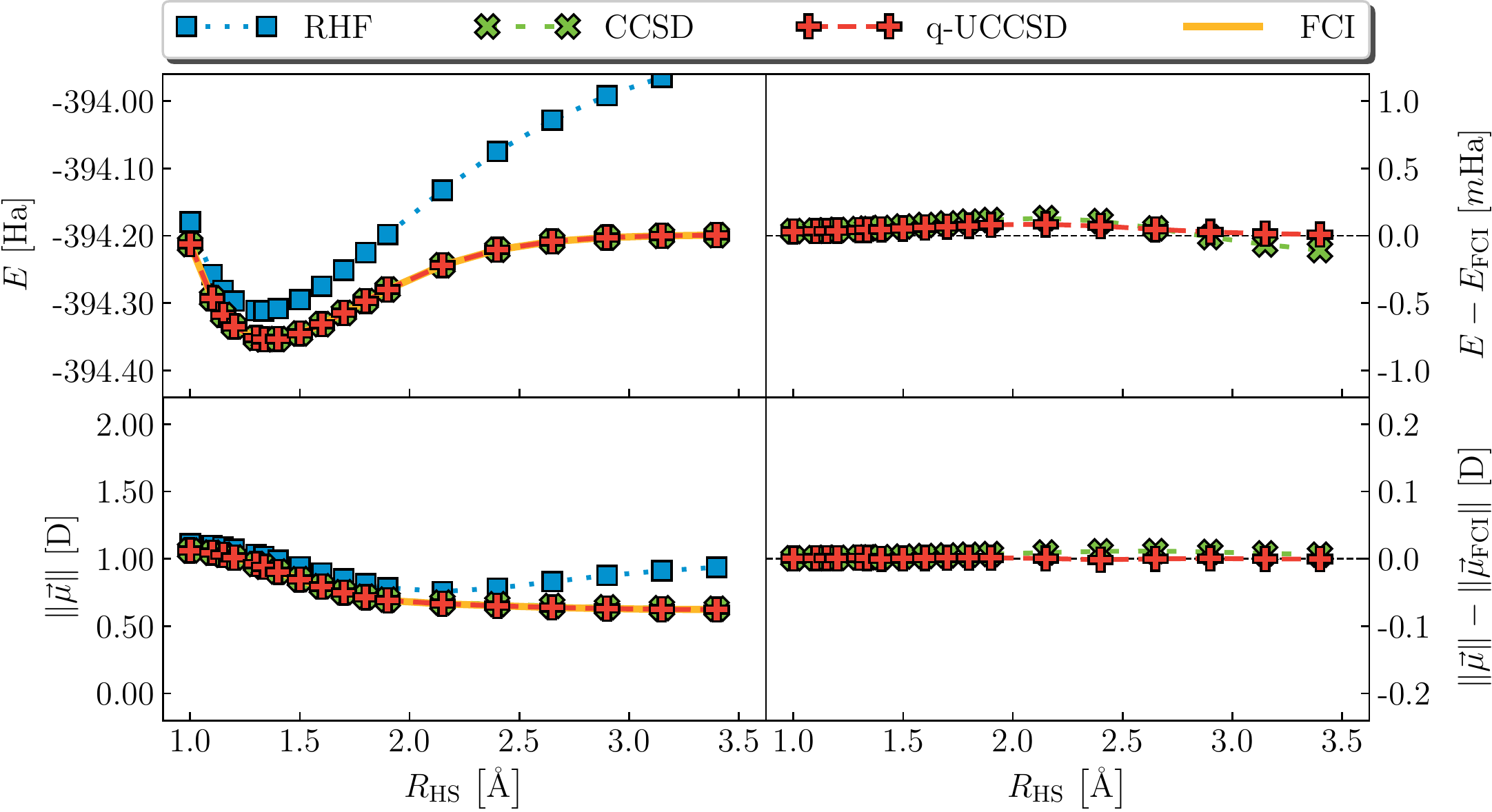}
\caption{
Calculations for \ce{H2S}.
Top left: dissociation curve as a function of HS distance from RHF, CCSD, q-UCCSD and FCI. 
Top right: Deviation from FCI energy, for CCSD and q-UCCSD. 
Bottom left: norm of the dipole moment as a function of HS distance.
Bottom right: deviation from FCI dipole, for CCSD and q-UCCSD.
}
\label{figure:3}
\end{figure*}
\begin{figure*}[t!]
\includegraphics[width=0.6\textwidth]{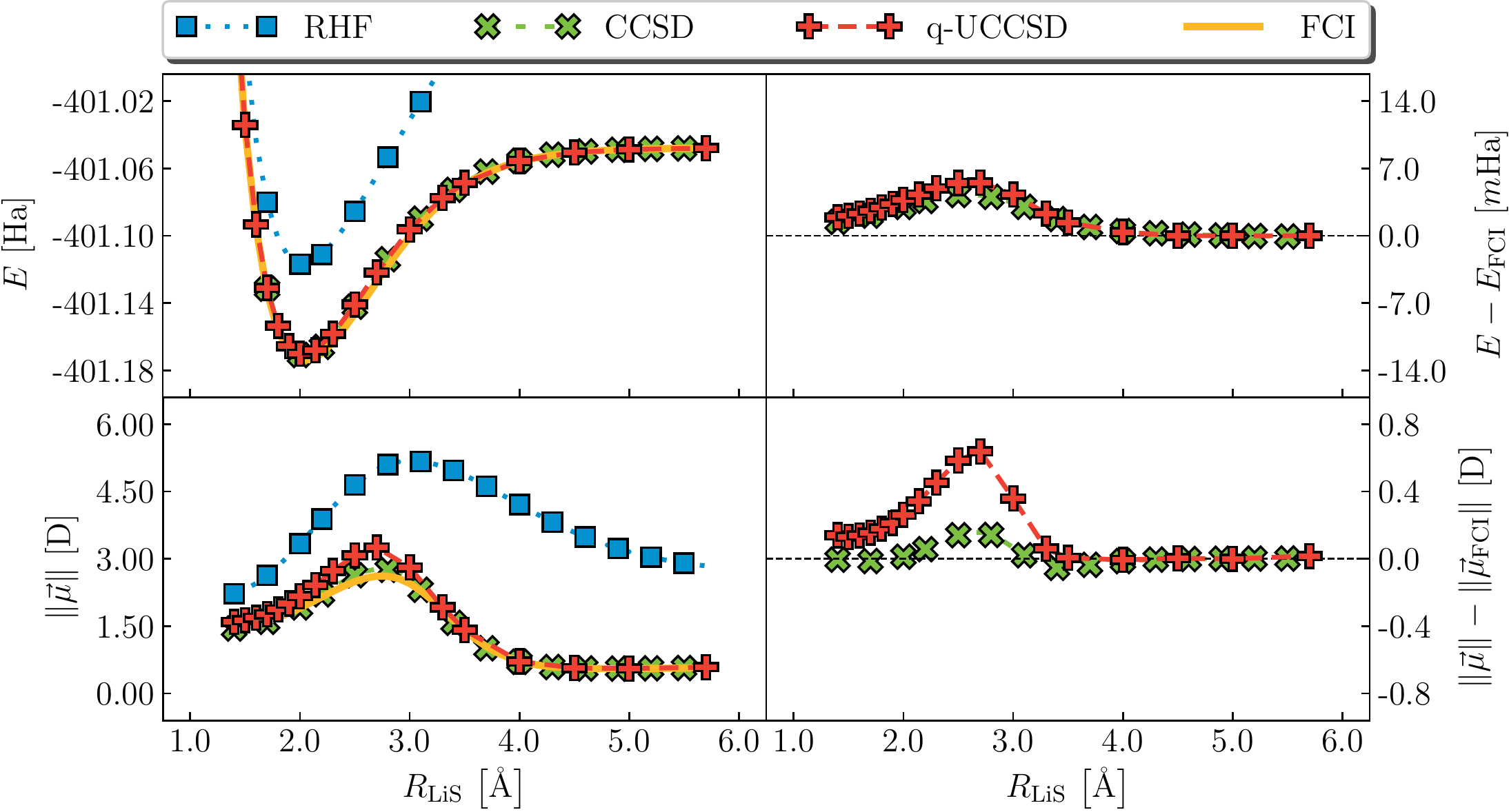}
\caption{
Calculations for \ce{LiSH}, when breaking the \ce{Li}-\ce{S} bond. 
Top left: dissociation curve as a function of LiS distance from RHF, CCSD, q-UCCSD and FCI. 
Top right: Deviation from FCI energy, for CCSD and q-UCCSD. 
Bottom left: norm of the dipole moment as a function of LiS distance.
Bottom right: deviation from FCI dipole, for CCSD and q-UCCSD.
}
\label{figure:4}
\end{figure*}
\begin{figure*}[t!]
\includegraphics[width=0.6\textwidth]{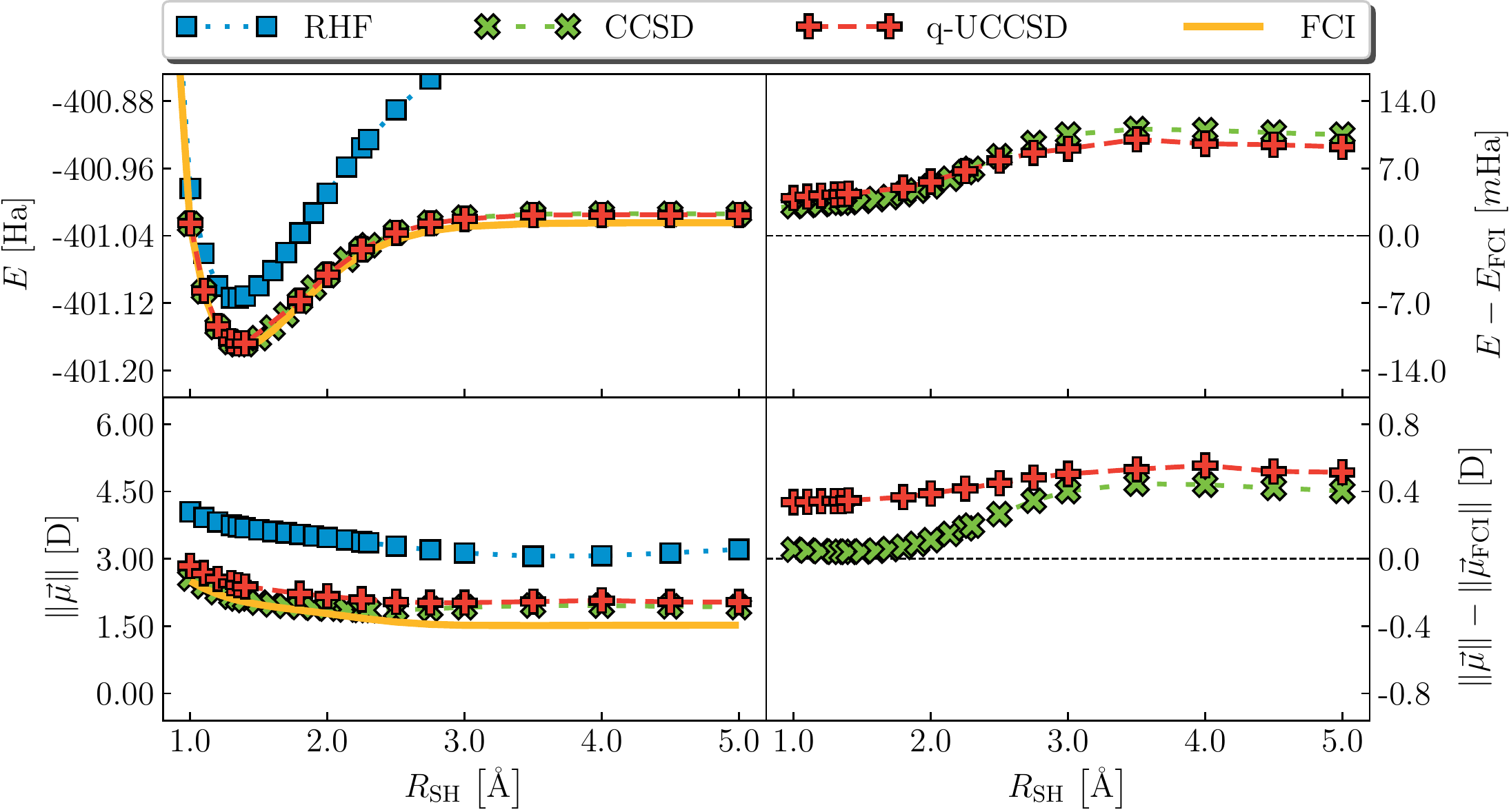}
\caption{
Calculations for \ce{LiSH}, when breaking the \ce{H}-\ce{S} bond. 
Top left: dissociation curve as a function of HS distance from RHF, CCSD, q-UCCSD and FCI. 
Top right: Deviation from FCI energy, for CCSD and q-UCCSD. 
Bottom left: norm of the dipole moment as a function of HS distance.
Bottom right: deviation from FCI dipole, for CCSD and q-UCCSD.
}
\label{figure:5}
\end{figure*}

\begin{figure*}[t!]
\includegraphics[width=0.6\textwidth]{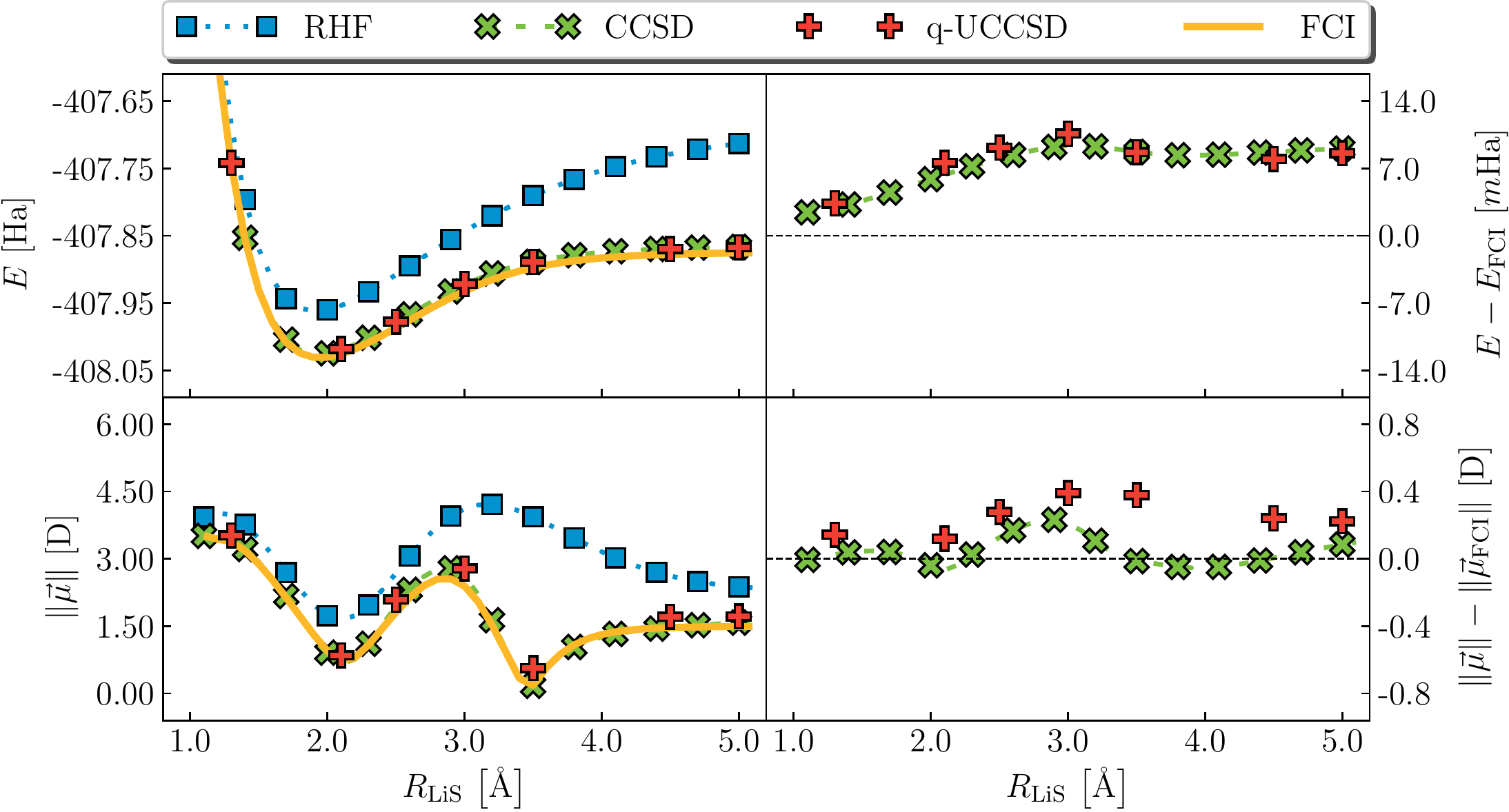}
\caption{
Calculations for lithium sulfide, \ce{Li2S}.
Top left: dissociation curve as a function of LiS distance from RHF, CCSD, q-UCCSD and FCI. 
Top right: deviation from FCI energy, for CCSD and q-UCCSD. 
Bottom left: norm of the dipole moment as a function of LiS distance.
Bottom right: deviation from FCI dipole, for CCSD and q-UCCSD.
}
\label{figure:6}
\end{figure*}

\subsubsection{Hydrogen sulfide $(\ce{H_2S})$}

After removing frozen core orbitals, hydrogen sulfide (\ce{H2S}) is represented by MOs with 3s, 
3p$_x$, 3p$_y$, 3p$_z$ character for sulfur and an MO with 1s character for hydrogen, totaling 
12 spin orbitals and 8 electrons. 
Upon employing tapering and associated qubit reductions to exploit symmetries in the molecule \cite{bravyi2017tapering,setia2019reducing}, the q-UCCSD ground state energy and dipole moment norm
were calculated over a range of bond distances on the simulator with 9 qubits as illustrated 
in Figure \ref{figure:3}. Also plotted are the corresponding CCSD and FCI energies and dipole moment norms.

q-UCCSD and CCSD are in agreement with the FCI energies across bond stretching 
(the maximum deviation in milliHartrees from the FCI value being 0.085 (at 2.15 \AA) and 0.131 
(at 2.15 \AA) respectively,
with q-UCCSD providing a better estimate of the energy for large $R_{\ce{HS}}$. 
Note that the the CCSD energy actually goes below the FCI value at large $R_{\ce{HS}}$, 
a known deficiency of CCSD to describe the molecular energy when the bond length is far from the equilibrium value.

The dipole moment norm is seen to monotonically decrease with $R_{\ce{HS}}$ towards $\sim$ 0.67 D, 
which is the dipole moment norm of the HS fragment, and again q-UCCSD values are in better 
agreement with FCI than CCSD ones 
(the maximum deviations from FCI being 2 milliDebye and 12 milliDebye respectively).

As seen in Figure \ref{figure:3}, differences between q-UCCSD and FCI energies are always greater 
than zero. This can be understood, since the q-UCCSD energies are variational in nature \cite{kutzelnigg1982quantum,kutzelnigg1983quantum,kutzelnigg1985quantum}
but since the method only includes single and double excitations, the values must lie above the 
FCI values. 
It is also understandable that the most challenging regime is the intermediate dissociation 
regime ($R_{\ce{HS}} \simeq 1.8$ \AA) where the ground-state wavefunction is switching between 
different dominant determinants in its configuration interaction expansion, exhibiting 
multi-reference character.

$ $ 

$ $

$ $

$ $

$ $

$ $

$ $

\subsubsection{Lithium hydrogen sulfide $(\ce{LiSH})$}

After removing core orbitals, lithium hydrogen sulfide (\ce {LiSH}) is represented by MOs 
with 3s, 3p$_x$, 3p$_y$, and 3p$_z$ 
character for sulfur, 1s character for hydrogen and 2s, 2p$_x$, 2p$_y$, 2p$_z$, 
character for lithium, for a total of 18 spin
orbitals and 8 electrons. After employing tapering and qubit reductions possible due to symmetry \cite{bravyi2017tapering,setia2019reducing}, the ground state energy and dipole moment norms were calculated on
the simulator with 15 qubits.

Results are shown in Figures \ref{figure:4} and \ref{figure:5}. 
In Figure \ref{figure:4}, the dissociation of the \ce{LiS} bond is studied by varying $R_{\ce{LiS}}$ 
with all other internal coordinates fixed at their experimental values\cite{johnson1999nist}. 
For all but the largest values of $R_{\ce{LiS}}$, CCSD is in better agreement with FCI for both the energy and the dipole moment norm. 
As in the \ce{H2S} case, the most challenging regime is the intermediate dissociation region, where the q-UCCSD
energy is overestimated by 5 milliHartrees and the dipole moment norm by $\sim$20\% at 
$R_{\ce{LiS}} \simeq 2.7$ \AA.
Note that, just as for \ce{H_2S}, at large $R_{\ce{LiS}}$, the dipole moment norm converges towards $\| \vec{\mu} \| \simeq$ 0.67 D, the dipole moment of the HS fragment.

A very different situation is seen for the dissociation of the SH bond, illustrated in Figure \ref{figure:5}. 
Description of the dissociation regime is more challenging for both CC flavors than that seen for the breaking of the \ce{Li}-\ce{S} bond, 
with q-UCCSD again performing better at large bond length. The FCI dipole moment norm decreases monotonically with $R_{\ce{SH}}$, 
a trend that both CC flavors reproduce qualitatively, but not quantitatively, particularly at longer bond lengths. 
Note that, in the large $R_{\ce{SH}}$ regime, CCSD and q-UCCSD dipole norms converge towards $\| \vec{\mu} \| \simeq$ 2.10 D, which is higher than the FCI value (1.52 D), signaling the limited accuracy of the underlying Ans\"atze.

\begin{figure}[b!]
\includegraphics[width=0.8\columnwidth]{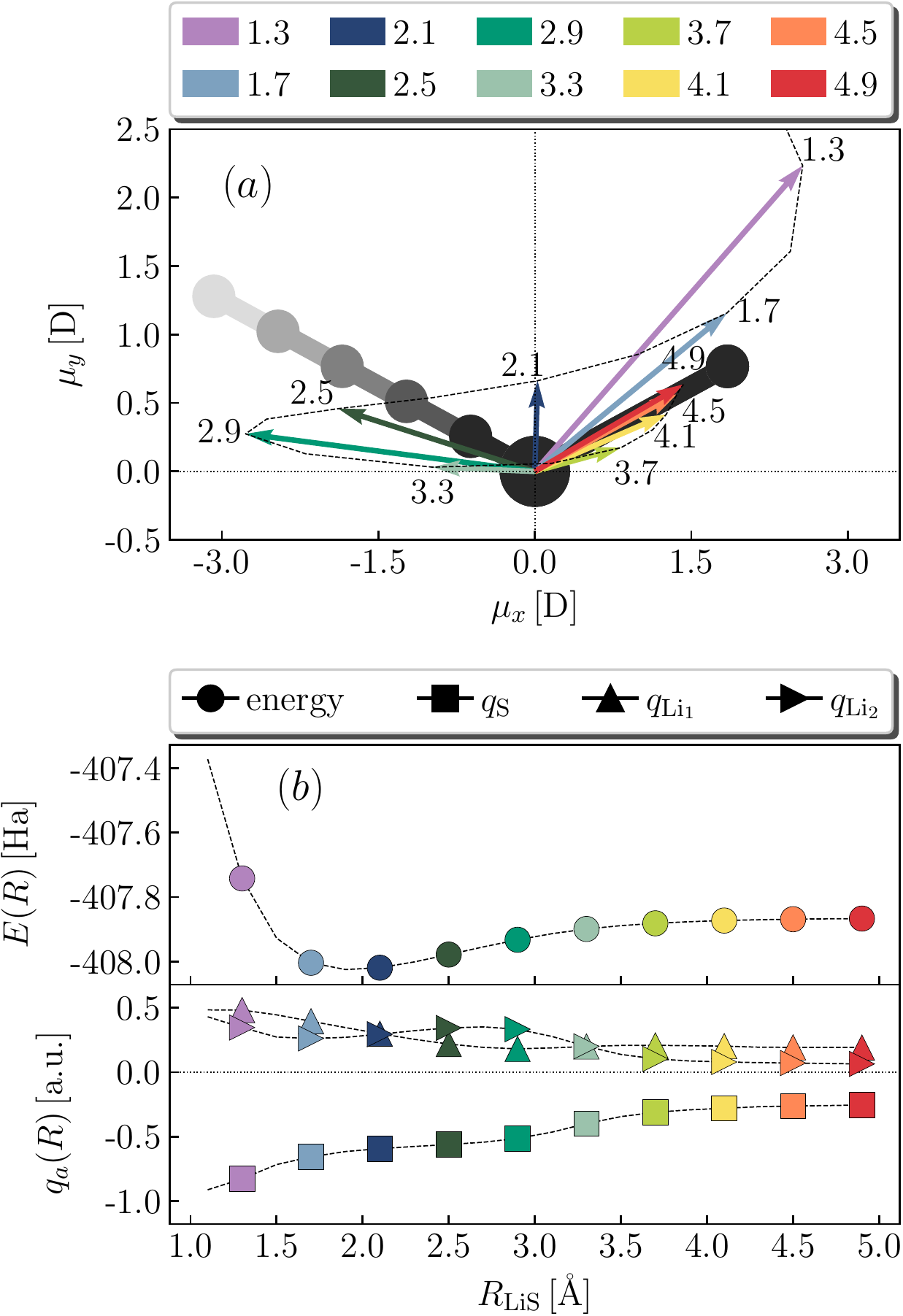}
\caption{(a) Evolution of the dipole moment vector of LiS along breaking of the LiS bond. The length of the arrow is the dipole moment norm, shown in Figure \ref{figure:6}.
The black circles in the background denote the S (large) and fixed Li (small) atom, and the moving gray circles illustrate the departing Li atom.
(b) Evolution of energy (top) and partial charges (bottom) of LiS along breaking of the LiS bond.
Colors correspond to different bondlengths, dotted black lines are a guide for the eye, illustrating the evolution of dipole vector, energy and charges with bondlength.
}
\label{figure:7}
\end{figure}

\subsubsection{Lithium sulfide $(\ce{Li_2S})$}

\begin{figure*}[t!]
\includegraphics[width=0.65\textwidth]{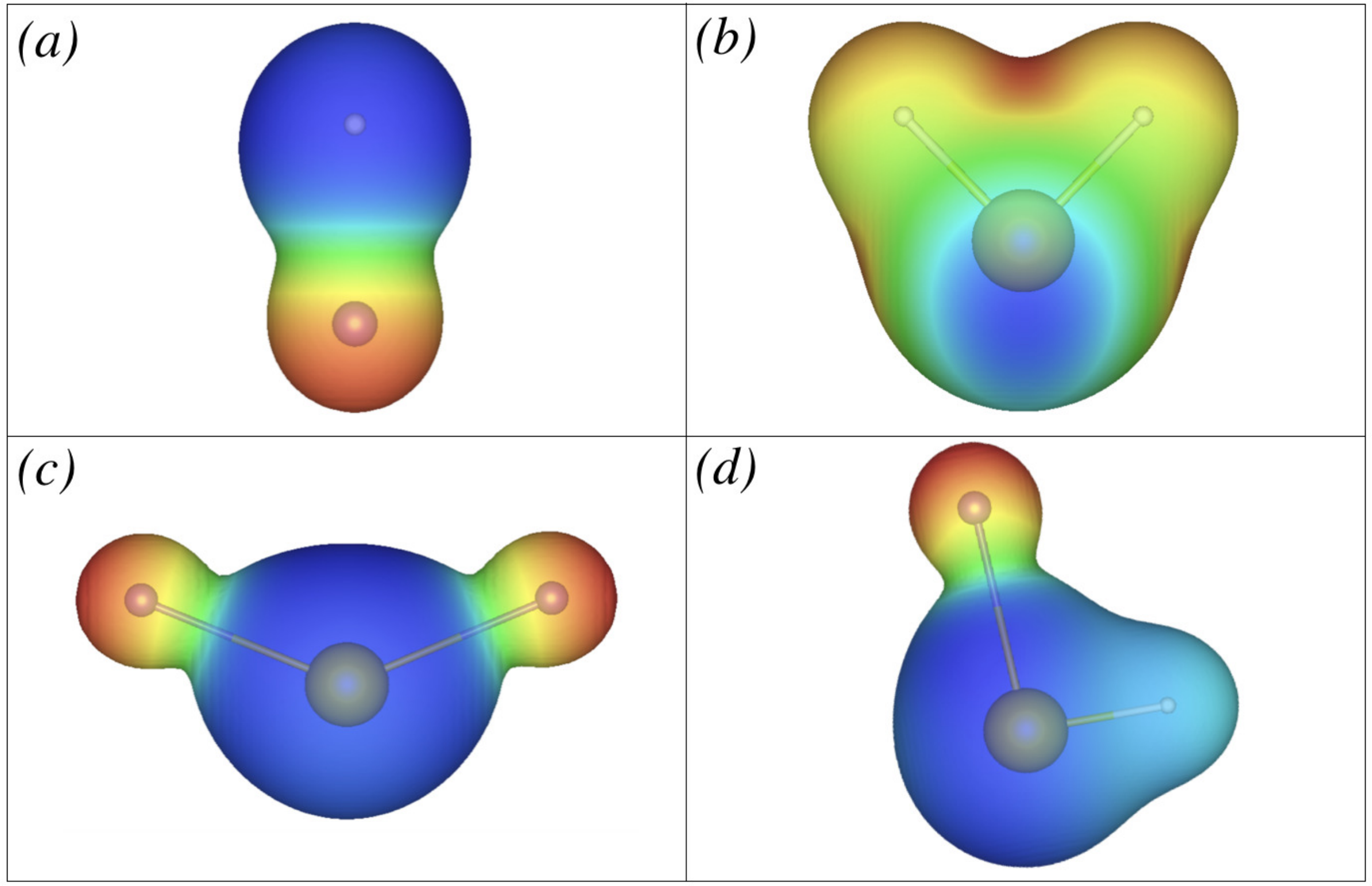}
\caption{Contour plots of the q-UCCSD molecular electrostatic potential
of LiH, H$_2$S, Li$_2$S and LiSH (a,b,c,d respectively) at equilibrium geometry.
Contour plots are shown along the isosurfaces of the q-UCCSD charge density with values $0.025$, $0.02$,
$0.025$ and $0.025$ respectively. Molecular electrostatic potential values range between minimum values -0.05, -0.04, -0.02, -0.04 (blue) and maximum values 0.60, 0.12, 0.53, 0.53 (red) respectively.
White, purple and yellow spheres represent H, Li and S atoms respectively.
}
\label{figure:8}
\end{figure*}

\begin{table*}
\begin{ruledtabular}
\begin{tabular}{ccrrrrrrrr}
& & \multicolumn{2}{c}{LiH} & \multicolumn{2}{c}{H$_2$S} & \multicolumn{2}{c}{LiSH} & \multicolumn{2}{c}{Li$_2$S} \\
\hline
component & unit      & q-UCCSD  & FCI     & q-UCCSD & FCI & q-UCCSD & FCI & q-UCCSD & FCI \\
\hline
$Q_{10}$  &     D     &  -1.8286 & -1.8286 & 0.3520 & 0.3518   & 0.9445  & 0.8079 & 0.3341 & 0.2853 \\
$Q_{11c}$ &     D     &          &         &        &          & 0.1159  & 0.1154 &        &        \\
\hline
$Q_{20}$  & D \AA     &  -5.0908 & -5.0908 & -8.8799 & -8.8805 & -6.0715 & -6.5529 & -15.3387 & -15.4943 \\
$Q_{21c}$ & D \AA     &          &         &         &         & -1.3555 & -1.3962 &          &          \\
$Q_{22c}$ & D \AA     &          &         &  0.7224 &  0.7224 & 1.4482  &  1.4910 & 8.8109   & 8.5606   \\
\hline
$Q_{30}$  & D \AA$^2$ & -17.6025 & -17.6026 & -3.0819 & -3.0828 & -20.3586 & -17.6287    & -11.1787 & -11.8992 \\
$Q_{31c}$ & D \AA$^2$ &          &          &         &         &  5.3521 &  5.5159 &         &         \\
$Q_{32c}$ & D \AA$^2$ &          &          & -0.2627 & -0.2634 & -3.5015 & -3.6918 & 17.3761 & 17.0057 \\
$Q_{33c}$ & D \AA$^2$ &          &          &         &         & -0.8122 & -0.8108 &          &        \\
$Q_{33s}$ & D \AA$^2$ &          &          &         &         & -0.0664 & -0.0661 &          &        \\
\hline
$Q_{40}$  & D \AA$^3$ & -77.1731 & -77.1730 & -48.7331 & -48.7366 & -91.8761 & -105.8862 &   -169.7058 & -173.6052 \\
$Q_{41c}$ & D \AA$^3$ &          &          &          &          & -6.5822 & -7.4223 & & \\
$Q_{42c}$ & D \AA$^3$ &          &          & -1.9550  & -1.9565  & 15.1501 & 16.1430 &   -26.5829 & -29.2577 \\
$Q_{43c}$ & D \AA$^3$ &          &          &          &          & -9.4300 & -9.3020 & & \\
$Q_{43s}$ & D \AA$^3$ &          &          &          &          & -0.3018 & -0.2849 & & \\
$Q_{44c}$ & D \AA$^3$ &          &          &  0.1150  &  0.1146 & 0.8223 & 0.8661&   68.6044 & 68.6344 \\
$Q_{44s}$ & D \AA$^3$ &          &          &          &          & -2.7937 & -2.8335 & & \\
\end{tabular}
\end{ruledtabular}
\caption{Multipole moments for LiH, H$_{2}$S, LiSH and Li$_2$S (top to bottom) in traceless spherical Buckingham form, at equilibrium geometry, in the center of mass frame. Blank entries correspond to multipole moments that are exactly zero due to the symmetry.
}\label{table:B1}
\end{table*}

\begin{figure*}[!t]
\includegraphics[width=0.75\textwidth]{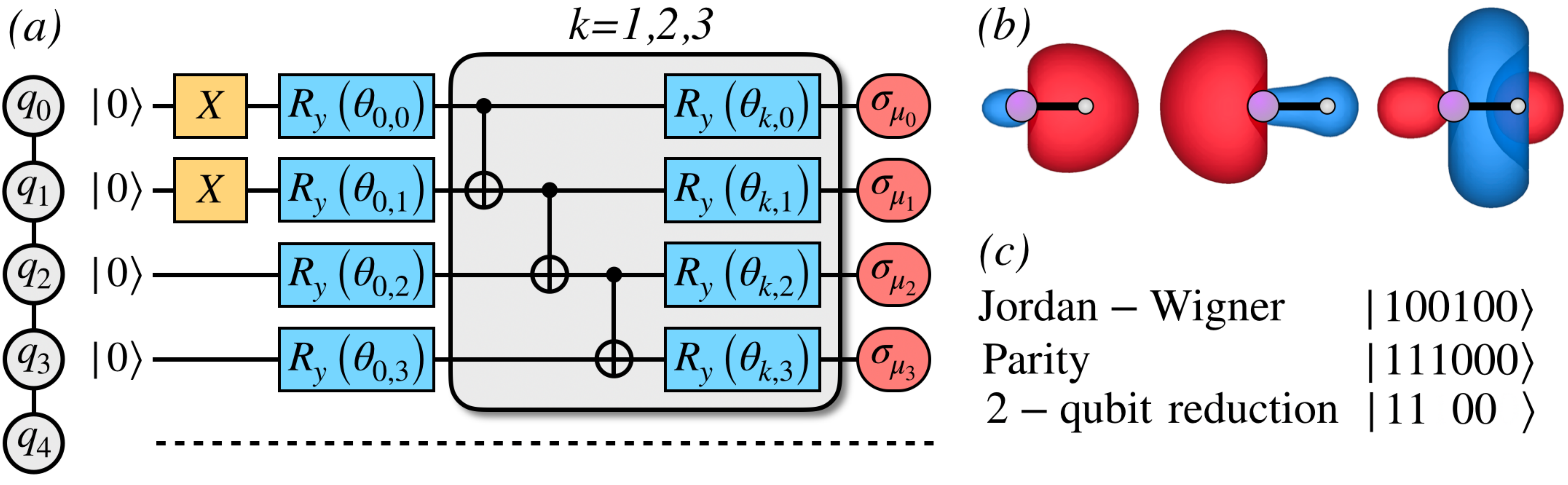}
\caption{(a) Schematic representation of the 5-qubit linearly-connected devices employed in this work, and of the quantum circuit used to simulate the ground-state energy and dipole moment of LiH. Orange symbols denote gates encoding of the Hartree-Fock state, blue symbols the $y$-rotations defining the variational Ansatz, and red symbols the measurements of Pauli operators.
(b) Molecular orbitals (MO) encoded on the quantum hardware, for LiH at equilibrium geometry, respectively the highest occupied MO and the two virtual MOs in the A$_1$ irreducible representation of the $C_{\infty v}$ group (left to right).
(c) 
Binary strings encoding the Hartree-Fock wavefunction under Jordan-Wigner and parity mappings, without and with removal of the two qubits related to the spin-up and spin-down particle number conservation in the parity mapping (top to bottom).
}
\label{figure:9}
\end{figure*}

After removal of the core orbitals for lithium and sulfur, lithium sulfide \ce{Li2S} is represented by MOs with 3s, 3p$_x$, 3p$_y$, 3p$_z$ character for sulfur 
and 2s, 2p$_x$, 2p$_y$, 2p$_z$ character for lithium, for a total of 24 spin orbitals and 8 electrons. 
After employing tapering and qubit reductions to account for molecular symmetries, 
the ground state energy and dipole moment norms were calculated on the simulator with 21 qubits. 
q-UCCSD and CCSD estimates for energy and dipole are similar in accuracy along bond stretching.

Results are given in Figure \ref{figure:6}. Interestingly, we observe that the norm of the dipole moment varies
non-trivially with $R_{\ce{LiS}}$ for both CCSD, q-UCCSD, and FCI: 
it decreases up to 
$R_{\ce{LiS}} \simeq 2.1$ \AA, 
then increases until $R_{\ce{LiS}} \simeq 3.0 $ \AA, 
then decreases until $R_{\ce{LiS}} \simeq 3.5 $ \AA, and then increases again 
towards an asymptotic value.
For q-UCCSD and FCI, this asymptotic value is of $\| \vec{\mu}\| \simeq 1.72$ D and $1.49$ D respectively.

This behavior is elucidated in Figure \ref{figure:7}, where we consider the evolution of
the dipole vector of $\ce{Li_2S}$ along breaking of a single LiS bond. As the bond length evolves from 1.3 to 2.1 \AA $\,$ (equilibrium), the dipole moment decreases in length, and rotates towards the bisector of the Li-S-Li triangle.

As the bondlength further increases, the dipole moment increases again in length, rotating towards the departing Li atom, until the point of inflection of the potential energy curve ($R_{\ce{LiS}} \simeq 3.0$ \AA)
is reached.
There, at q-UCCSD level, it suddenly rotates towards the unbroken LiS bond. Note that FCI values for the dipole moment of LiSH at long SH distance are very close to those of $\ce{Li_2S}$ at long LiS distance (1.52 and 1.49 Debye respectively), since they both result in a fragment of LiS.

\subsection {Study of electrostatic properties}

In this section, we evaluate multipole moments, 
charge densities and electrostatic potentials
for the species discussed in the previous section,
focusing on equilibrium geometries.

In Figure \ref{figure:8}, we evaluate charge densities and molecular electrostatic potentials.
The contour plots  qualitatively illustrate the electronegative nature of sulfur, 
which hosts negative partial charges in compounds (b,c,d). On the other hand, as expected,
lithium has low electronegativity and hosts positive partial charges in compounds (a,c,d).

In Table \ref{table:B1}, 
we list the multipole moments for LiH, 
H$_2$S, LiSH and 
Li$_2$S respectively.
The comparison between q-UCCSD and FCI demonstrate the accuracy of the former:
higher multipole moments, which are notoriously very sensitive physical quantities to the choice of method, exhibit
somewhat large deviations. 
Such a limited accuracy is also a reflection of the heuristic nature of the q-UCCSD Ansatz,
and pinpoints the need of developing more accurate Ans\"atze and methodologies, in order to satisfactorily
describe sensitive electrostatic properties of molecules.

We note that molecular electrostatic potentials can be used to obtain partial charges through the restrained electrostatic potential (RESP) technique
\cite{resp1,resp2,resp3,resp4},
offering further insight into 
the interaction between molecules and their dissociation. For example, in Figure \ref{figure:7}, we show the partial charges of $\ce{Li_2S}$ along breaking of the LiS bond: the
Li atom departing from the molecule is neutral in the large $R_{\ce{LiS}}$ limit, while the remaining Li and S atoms
carry partial charges responsible for the asymptotic value of the dipole moment.

\begin{figure}[!h]
\includegraphics[width=0.85\columnwidth]{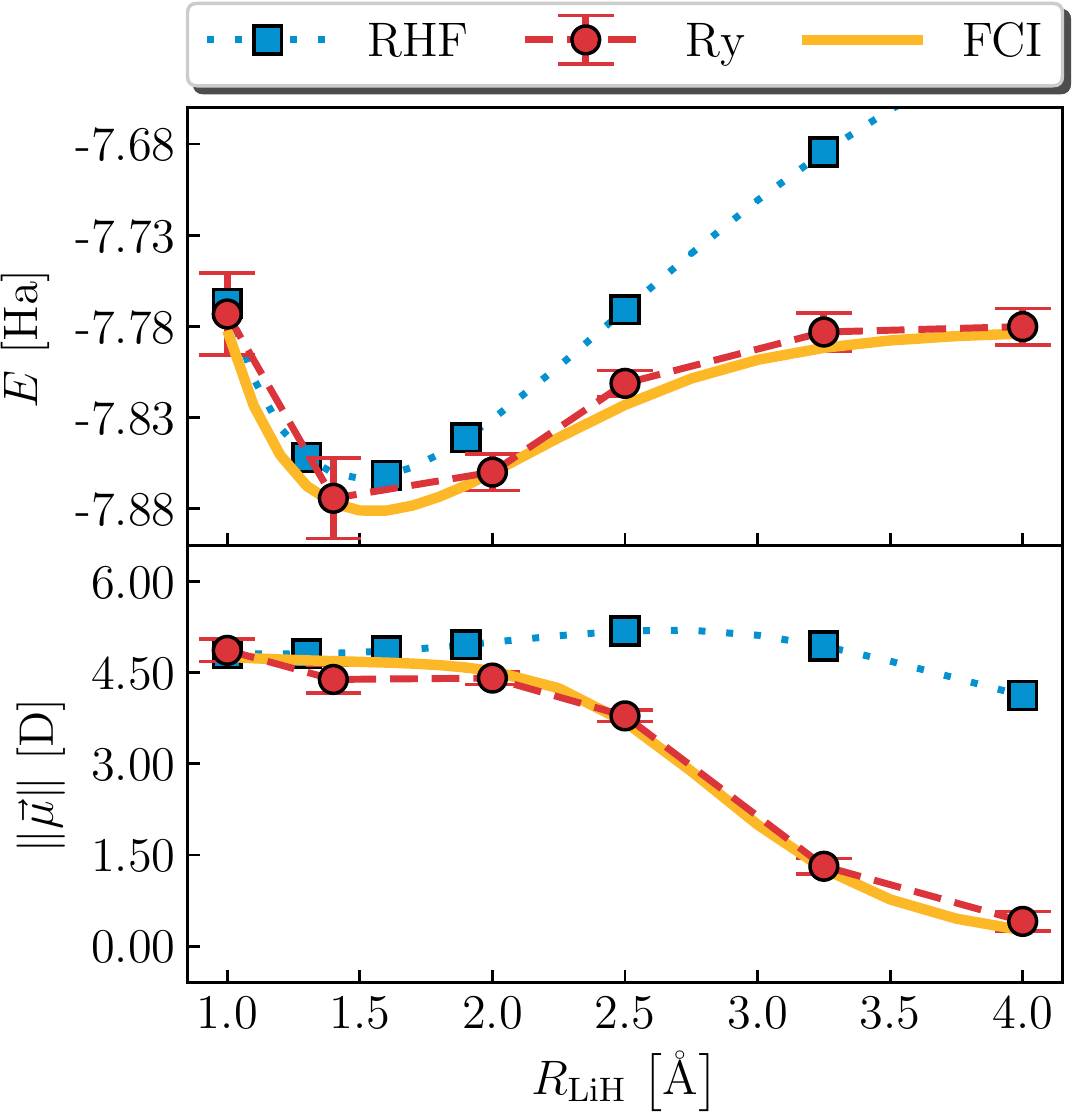}
\caption{Hardware results of ground-state energy (top) and dipole moment  (bottom) calculations for \ce{LiH} on the 5 qubit devices accessed via IBM Quantum Experience. (Top: energy in Hartrees as a function of interatomic distance, in Angstroms.}
\label{figure:10}
\end{figure}

\subsection{Quantum chemistry calculations on quantum devices}

Next, the dipole moment of lithium hydride was evaluated using quantum hardware as sketched in Figure \ref{figure:9}. 
We employ the hardware-efficient $\mathrm{R_y}$ Ansatz to estimate the ground-state energy and dipole moment at representative 
values of $R_{\ce{LiH}}$. We used a circuit with $\mathsf{cNOT}$ entangling gates with linear connectivity and
3 layers
as illustrated in Figure \ref{figure:9}. 

To the best of our knowledge, these are the first evaluations of dipole moments on quantum hardware. 

We chose to study six bond lengths representative of the polar and dissociation regimes. 
We chose the $\mathrm{R_y}$ Ansatz because experience \cite{kandala2017hardware} suggests that, for small molecules like LiH, it can deliver accurate ground state energies with modest quantum resources, unlike q-UCCSD.

Results from hardware experiments in presence of readout error mitigation are shown. A Richardson extrapolation \cite{temme2017error,kandala2019error} 
is also conducted, with the aim of further mitigating the impact of noise (intermediate data are listed in the Supplementary Material). 
As seen in Figure \ref{figure:10}, the qualitative behavior of both energies and dipoles is correctly captured
by the hardware experiments upon extrapolation. In particular, the dipole moment is accurate within $\sim$10\% in the polar regime. 
Future research will explore approaches to improve these results quantitatively, for example by the use of circuits with higher depth, 
by the exploration of multiple Ans\"{a}tze or by the use of different error mitigation techniques.

\section*{Conclusions}

Herein, we have reported on the simulations of four molecules 
(\ce{LiH}, \ce{H2S}, \ce{LiSH} and \ce{Li2S}) that increase in complexity to \ce{Li2S} 
which is relevant to the study of lithium-sulfur batteries. 
We showed that we could obtain ground state energies for LiH and $\ce{H_2S}$
within 1 mHa from FCI. For species with more valence electrons, our calculations pinpoint the limitations of Ans\"atze based on single and double electronic excitations, especially away from equilibrium.

Differences could be observed in the \ce{LiSH} molecule, depending on whether the 
hydrogen or lithium atom was dissociated from the sulfur atom: 
when the \ce{SH} bond is dissociated, the dipole moment decreases monotonically 
towards a large ($\sim$ 1.5 D) asymptotic limit, and when the \ce{LiS} bond is dissociated
it converges non-monotonically towards a smaller asymptotic value (0.57 D).
The asymptotic values of the dipole moment norms are a reflection of the greater polarity
of the \ce{LiS} bond compared to the \ce{SH} bond.
The non-monotonicity, on the other hand, is analogous to the behavior seen in the 
\ce{Li_2S} molecule, which is due to non-trivial changes in the direction of the 
dipole moment as $R_{\ce{LiS}}$ changes.

Furthermore, the effects of electronic distribution changes could be observed when 
evaluating the dipole moment of \ce{LiH}. Additionally, we showed that the \ce{LiH}
could be qualitatively determined on quantum hardware with 4 qubits. 
This is a notable demonstration of the capabilities of the hardware since the dipole 
moment of lithium hydride changes from strongly polarized in its ionic state to 
effectively neutral at long bond distances over $\sim$ 2.5 \AA, representing a highly 
entangled state at this distance. 

Characterizing the electronic structure of Li-S compounds is an important step towards understanding lithium-sulfur batteries. However, achieving this goal with quantum computers requires
further methodological improvements. For example, achieving the ability to tackle larger and more relevant chemical species \cite{manthiram2014rechargeable,chung2018progress,li2019comprehensive} and non-minimal basis sets \cite{dunning1989gaussian,woon1993gaussian,dunning2001gaussian}, developing Ans\"{a}tze that achieve a balance between hardware efficiency and chemical insights \cite{grimsley2019adaptive,mccaskey2019quantum,takeshita2020increasing,motta2020quantum,nam2020ground}, and incorporating nuclear motion effects \cite{ollitrault2020hardware,stober2020computing,sawaya2020resource,sawaya2020near}.

Our work provides a stepping stone on the way to larger quantum computing calculations, on polyanions formed upon discharge in the Li-S battery. These calculations will have the potential to compare between radical and ionic mechanisms of the electrochemical reduction of sulfur to lithium sulfide 
with lithium metal.

\section*{Supplementary Material}

See supplementary material for details of both the quantum experiments and the  classical simulations, including investigation of the \ce{LiH} CCSD energy and dipole moment across the dissociation profile using different basis sets.

\section*{Acknowledgments}

We are indebted to many colleagues for helpful discussions, particularly Stephen Wood, Panagiotis Barkoutsos and Ivano Tavernelli.

TG, JL, MM and JER acknowledge the IBM Research Cognitive Computing Cluster service for providing resources that have contributed to the research results reported in this paper.

This paper is dedicated to the memory of Prof. Dr. Dr. Andreas Hintennach.

\section*{Data Availability}

The data that support the findings of this study are available from the corresponding author upon reasonable request.



%

\end{document}